\documentclass[prl,twocolumn,showpacs,a4paper]{revtex4}
\usepackage{graphicx}
\usepackage{amsmath}
\begin{document}

\title{Interaction-controlled transport of an ultracold Fermi gas}

\author{%
Niels Strohmaier$^1$, Yosuke Takasu$^{1, 2}$, Kenneth G\"unter$^1$, Robert
J\"ordens$^1$,\\
Michael K\"ohl$^{1, 3}$, Henning Moritz$^{1, *}$, and Tilman
Esslinger$^1$}

\affiliation{%
$^1$Institute for Quantum Electronics, ETH Zurich,
8093 Zurich, Switzerland\\
$^2$Department for Electronic Science and Engineering,
Kyoto University, Kyoto 615-8510, Japan\\
$^3$Cavendish Laboratory, University of Cambridge,
Cambridge CB3 0HE, United Kingdom}

\date{\today}

\begin{abstract}
We explore the transport properties of an interacting Fermi gas in a
three-dimensional optical lattice. The center of mass dynamics of the atoms
after a sudden displacement of the trap minimum is monitored for different
interaction strengths and lattice fillings. With increasingly strong
attractive interactions the weakly damped oscillation, observed for the
non-interacting case, turns into a slow relaxational drift. Tuning the
interaction strength during the evolution allows us to dynamically control
the transport behavior. Strong attraction between the atoms leads to the
formation of local pairs with a reduced tunneling rate. The interpretation
in terms of pair formation is supported by a measurement of the number of
doubly occupied lattice sites. This quantity also allows us to determine the
temperature of the non-interacting gas in the lattice to be as low as
$(27\pm2)\%$ of the Fermi temperature.

\end{abstract}


\maketitle

The study of conductivity in solids has led
to the discovery of fundamental phenomena in condensed matter
physics and to a wealth of
knowledge on electronic phases.
Intriguing quantum many-body phenomena such as superconductivity and
the quantum Hall effect manifest themselves in their characteristic
electronic transport properties.
Moreover, the ability to manipulate
conductivity has found numerous applications in technology, most
prominently in semiconductors, where transport is controlled by electric
biasing. Other ways to modify the conductivity in a material include
adjusting temperature, pressure or magnetic field.

A gas of ultracold fermionic atoms exposed to the potential of an optical
lattice offers a new approach to study and control transport while providing
a direct link to fundamental models in condensed matter physics. A periodic
potential of simple cubic symmetry is generated by three mutually
perpendicular laser standing waves reproducing the potential experienced by
electrons in the crystal structure of a solid. Prepared in two different
spin states, fermionic atoms mimic spin-up and spin-down electrons. A unique
feature of the atomic system is that the strength of the collisional
interaction between the two components can be directly tuned using a
Feshbach resonance \cite{Inouye1998, Loftus2002}. While this property has
been used to study fermionic superfluidity in the strongly interacting
regime (e.\,g.~\cite{BCSBEC,Chin2006}), it has so far not been applied to
investigate transport phenomena in optical lattices. In previous experiments
the transport of non-interacting fermionic atoms and the effect of a bosonic
admixture mediating interactions were studied in one-dimensional optical
lattices \cite{Ott2004a,Pezze2004,Ponomarev2006a}. Furthermore, the dynamics
of Bose gases in a three-dimensional optical lattice was investigated
experimentally and theoretically
\cite{Fertig2005,Mun2007,TheoryTransport3DMI}.

\begin{figure}[htb]
  \includegraphics[width=.8\columnwidth]{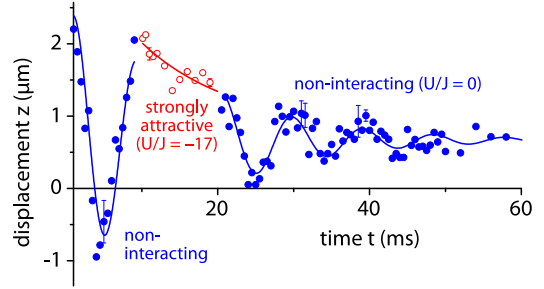}
  \caption{(Color online)
  Dynamic control of transport by tuning the collisional interaction.
  The graph shows
  the center of mass motion of a two-component Fermi gas of $(2.9\pm0.3)\times10^4$ atoms in a lattice of $5\,E_R$ depth.
  At $t=0$, the equilibrium position of the underlying harmonic trap
  is displaced vertically. After 9\,ms of evolution without interaction
  ($\bullet$), the magnetic field is changed linearly in 1\,ms
  so that the interaction is strongly attractive ($\circ$) for
  10\,ms. Then the magnetic field is changed to its original value again
  within $0.5$\,ms ($\bullet$).
  The two non-interacting cases $\bullet$ are each fit by a damped cosine
  and an offset, while the evolution with attractive interaction $\circ$ is
  fit by an exponential decay and an offset.
  Error bars denote the standard deviation of at least 4 measurements.}
  \label{fig1}
\end{figure}

In this letter we study the transport properties of a two-component $^{40}$K
cloud trapped inside a three-dimensional optical lattice with underlying
harmonic confinement. We monitor the center of mass motion of the atomic
cloud after a sudden displacement of the trap minimum. The regimes of
vanishing, weakly attractive and strongly attractive interactions are
accessed by exploiting a Feshbach resonance to tune the scattering length
for low energy collisions between the two atomic components. The atom number
is adjusted so that at the trap center the lowest energy band is either
filled or half-filled. For these parameters the system can be regarded as a
realization of the attractive single-band Fermi-Hubbard model
\cite{Micnas1990} with additional harmonic confinement. A Mott insulating
phase of pairs as discussed in the context of the multi-band Hubbard model
\cite{Chin2006,MIofPairs} is not expected.

The strong influence of the interactions on the transport is
illustrated in Fig.\,\ref{fig1}. An atomic cloud is prepared in the optical lattice at half
filling and brought into non-equilibrium by displacing the trap
minimum. The initially non-interacting cloud performs a weakly
damped oscillatory motion in the confining potential. By
temporarily switching on the attractive interaction, a controlled interruption of this oscillation is
achieved.

Our experimental setup that is used to produce quantum degenerate Fermi gases is
described in detail in previous work \cite{Koehl2005}. In brief, we prepare a cloud of
$^{40}$K atoms in an equal mixture of the hyperfine substates $|F=9/2, m_F=-9/2\rangle$ and $|F=9/2,
m_F=-7/2\rangle$ in a crossed-beam optical dipole trap
operating at a wavelength of 826\,nm. After evaporative cooling we obtain $4
\times10^4$ ($3\times10^5$) atoms at temperatures below $T/T_F=0.20~
(0.25)$
in the dipole trap with final trapping frequencies of $(\omega_x,\omega_y,
\omega_z)=2\pi\times(35,23,120)$\,Hz, where $T_F$ is the Fermi temperature.
Next, the degenerate Fermi gas is subjected to the additional periodic
potential of a three-dimensional optical lattice with a depth of $5\,E_R$.
The recoil energy is given by $E_R=h^2/(2 m\lambda^2)$, with $h$ being the
Planck constant, $m$ the atomic mass and $\lambda=1064$\,nm the wavelength
of the lattice beams. The lattice is formed by three retro-reflected laser
beams with circular profiles having $1/\mathrm{e}^2$ radii along the
$(x,y,z)$-directions of $(160,180,160)\,\mu$m at the positions of the atoms
and a mutual frequency difference of several 10\,MHz. To load the atoms into
the lowest Bloch band of the optical lattice we increase the intensity of
the lattice beams using a spline ramp with a duration of 100\,ms at a
scattering length of $a=50\,a_0$, where $a_0$ is the Bohr radius.

The gas is brought into a non-equilibrium position by increasing the beam
intensities of the underlying dipole trap, which shifts the trap minimum by
up to $2.5\,\mu$m in the vertical $z$-direction. Since this displacement is
smaller than our imaging resolution, we map the center of mass position of
the atomic cloud to momentum space. For this purpose we switch off the
optical lattice and let the cloud oscillate in the remaining harmonic dipole
trap for a quarter period \cite{Fertig2005}. After free expansion, we obtain
the momentum distribution of the cloud from absorption imaging, determine
the center of mass momentum using a Gaussian fit and infer the original
displacement $z$ of the cloud in the trap. Oscillations of the cloud size
are not observed since the horizontal and vertical trapping frequencies are
only increased by about 4\% and 12\%, respectively. Also, Bloch oscillations
can be neglected for our small displacement since even for large fillings
only few atoms gain sufficient energy to reach the band edge
\cite{Snoek2007}. The energy deposited in the system by the trap
displacement is estimated to increase the temperature in the lattice by an
amount of 0.05\,$T_F$.

Variation of the magnetic bias field in the vicinity of the Feshbach resonance
at 202.1\,G \cite{Loftus2002} allows us to tune the collisional
interaction between the two components of the Fermi gas.
Prior to the displacement of the trap the magnetic field is gradually ramped to final
values between 210\,G and 202.95\,G within 50\,ms, yielding an $s$-wave
scattering length ranging from 0 to $-1500\,a_0$. Using the description of a Hubbard model for cold atoms \cite{Jaksch98,Hofstetter2002}, this corresponds to an
effective interaction strength $U/J$ between 0 and $-24$. Here $U$ denotes
the on-site interaction energy of two atoms in a different spin
state, and $J$ is the matrix element for nearest-neighbor tunneling, which has a value of $J\approx h\times290$\,Hz for our lattice
depth.

\begin{figure}[htb]
  \includegraphics[width=.8\columnwidth,clip=true]{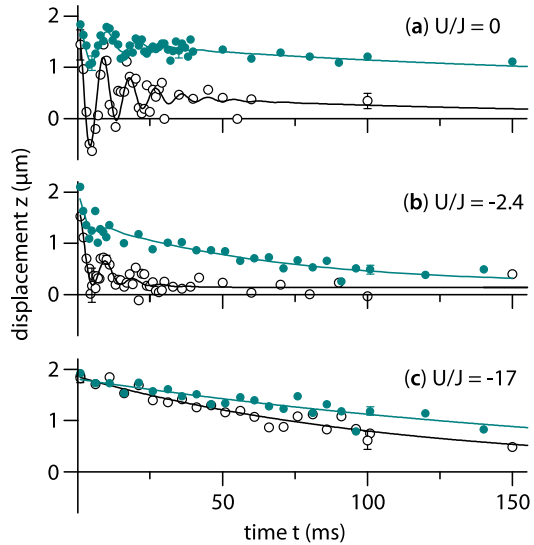}
   \caption{(Color online)
  Evolution of the center of mass position for different
  interaction strengths and fillings. The circles $\circ$ ($\bullet$)
  denote samples in the half filled (band insulating) regime.  For each
  data point the position of the cloud with and without displacement was
  compared to eliminate long term drifts. The error bars denote the
  standard deviation of at least 4 measurements.}
  \label{fig2}
\end{figure}

The results of transport measurements for three interaction strengths and two different fillings are shown in Fig.\,\ref{fig2}. We fit the function
$z(t)=z_\mathrm{osc}\, \cos(2\pi f t)\, \exp(-\beta t)+z_\mathrm{
exp}\,\exp(-\Gamma t) +z_0$ to the data.
For the half filled case,
$(3.6\pm0.4)\times10^4$ atoms are prepared corresponding to $0.46\pm0.05$ atoms per lattice site and spin state at the center of a
non-interacting cloud \cite{LatticeCalculations}. In the case of high
filling, the samples contain $(2.7\pm0.3)\times10^5$ atoms and form a large band insulating core. The filling factor of the Bloch band is position-dependent due to the inhomogeneous density profile of the trapped gas.
In the following, we will discuss the cases of zero, moderate and strong interaction.

In the non-interacting case with half filling we observe damped dipole
oscillations (Fig.\,\ref{fig2}(a)$\,\circ$). This damping of the center of
mass motion can be attributed to the fact that the fermions in different
quasi-momentum states possess different effective masses, resulting in a
spectrum of oscillation frequencies. Furthermore, the total trapping
potential is slightly anharmonic, which causes a dephasing also observed in
the pure dipole trap. The system with high filling
(Fig.\,\ref{fig2}(a)$\,\bullet$) is characterized by a very slow relaxation
towards the equilibrium position: The band insulating core suppresses center
of mass motion and a large number of atoms occupy localized states
\cite{Rigol2004,Ott2004b,Pezze2004}. These single particle eigenstates exist
at a distance $z_\mathrm{loc}$ from the center of the trap where the
potential energy due to the harmonic confinement is larger than the
bandwidth, i.\,e. $m \omega_z^2 z_\mathrm{loc}^2/2>4J$. Consequently, the
motion through the center is energetically prohibited, however, the atoms
can still oscillate within the outer regions of the cloud. Even in the half
filled case a small fraction of atoms is localized, which explains the small
offset observed in the center of mass position after the decay of the
oscillations (Fig.\,\ref{fig2}(a)$\,\circ$).

For moderate attractive interaction and half filling, the damping of the
dipole oscillations becomes more pronounced (Fig.\,\ref{fig2}(b)$\,\circ$).
The damping rate $\beta$ increases from $(80\pm 17)$ Hz in the
non-interacting case to $(140\pm 37)$ Hz at $U/J=-2.4$. As the interaction
strength is increased beyond $U/J<-3.5$, the oscillations vanish entirely.
The sample with high filling (Fig.\,\ref{fig2}(b)$\,\bullet$) relaxes faster
towards equilibrium than in the non-interacting case, which can be
attributed to umklapp processes \cite{Orso2004}.

\begin{figure}[htb]
  \includegraphics[width=.8\columnwidth,clip=true]{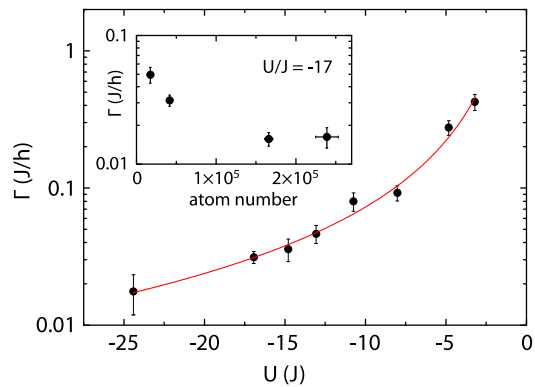}
  \caption{Relaxation rate $\Gamma$
  as a function of interaction strength. The data points and error bars
  are fit results to center of mass evolutions of $(3.8\pm0.4)\times10^4$
  atoms.
    The empirical power law $h\Gamma=C J (J/U)^\nu$ with
  $\nu=1.61\pm0.03$ and $C=2.95\pm0.17$ fits the data well.
    The inset shows the dependence of the relaxation rate on the
  atom number.}
  \label{fig3}
\end{figure}

In the strongly interacting case, a very slow relaxation is observed for
both fillings [Fig.\,\ref{fig2}(c)]. The transport in this regime is
governed by the dynamics of local fermionic pairs. In the limit of low
atomic densities bound pairs form for $U/J<-7.9$
\cite{Wouters2006,Micnas1990,Stoeferle2006}. These pairs tunnel to adjacent
sites via a second order process with an amplitude $J_\mathrm{eff}=2J^2/U$.
This effective tunneling is obtained by considering a ground state where all
atoms form pairs and by treating the tunneling term proportional to $J$ as a
perturbation in the Hubbard Hamiltonian \cite{Micnas1990}. Accordingly, the
tunneling rate of pairs is reduced with increasing interaction as compared
to bare atoms. Besides, the energy offset between neighboring sites due to
the harmonic confinement reduces the tunneling probability. For these
reasons we expect the relaxation time to become longer for stronger
interactions. This is supported by the data in Fig.\,\ref{fig3}, which shows
a clear decrease of the relaxation rate $\Gamma$ for growing attractive
interaction. The data is well fit by the empirical power law
$\Gamma/J\propto (U/J)^{-1.61}$. A quantitative understanding of this
behaviour is challenging due to the coexistence of bare atoms and local
pairs which act as hardcore bosons in the lattice.

\begin{figure}[htb]
  \includegraphics[width=.8\columnwidth,clip=true]{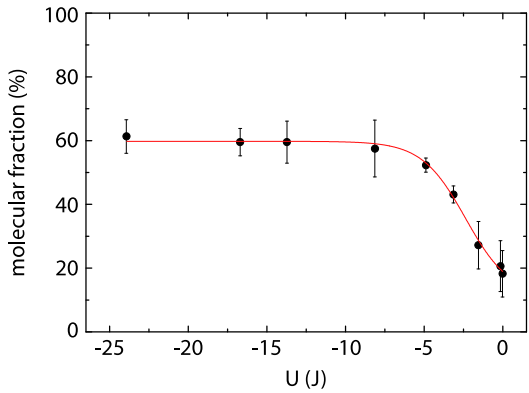}
  \caption{The fraction of molecules formed in the optical lattice increases
  with attractive interaction, demonstrating a higher number of doubly occupied lattice sites.
  Error bars denote the statistical errors of at least 4 measurements. The line serves as a guide to
the eye.}
  \label{fig4}
\end{figure}

Further insight into the physics of local pairs is gained by probing the double occupancy in the lattice for various interaction strengths without displacing the trap. For this purpose, we prepare the system at half filling, as before, and set the desired interaction within 50\,ms by changing the value of the magnetic field. Then the lattice depth is abruptly increased from 5\,$E_R$ to 30\,$E_R$ in order to suppress further tunnelling. By subsequently ramping the magnetic field from 203.26\,G to 201.23\,G within 5\,ms, weakly bound Feshbach molecules are formed on those sites which are doubly occupied \cite{Stoeferle2006}. We determine the number of atoms remaining after the molecule formation and compare it with the atom number which is obtained after dissociation of the molecules by applying the inverse magnetic field ramp. This yields the molecular fraction displayed in Fig.\,\ref{fig4}, showing a strong dependence on the interaction strength: While for the non-interacting system the detected fraction is 18\%, it increases up to 60\% for strongly attractive interactions.

For the non-interacting gas the double occupancy in the lattice is solely
determined by the number of trapped atoms and their temperature
\cite{Stoeferle2006,Koehl2006,LatticeCalculations}. The detected fraction of
18\% is consistent with the temperature in the lattice of
$(0.27\pm0.02)\,T_F$, which we determined in a separate measurement with
$2.7\times10^5$ atoms yielding a molecular fraction of $(45.7\pm2.4)\%$.
This temperature, even though measured in an ideal gas, suggests that the
gas remains above the critical temperature for superfluidity
\cite{Micnas1990,Keller2001,Santos94} also for strong interactions.
Numerical calculations for an homogeneous interacting system show a
considerable temperature dependence of the double occupancy
\cite{Keller2001}. We therefore expect that the temperature of the
interacting gas can be deduced from the measured double occupancy.

The increase of the molecular fraction with rising attractive interactions
provides strong evidence for the formation of local pairs. In accordance
with numerical calculations for the attractive Hubbard model at finite
temperature \cite{Keller2001,Santos94} the number of doubly occupied sites
increases already for weak attraction, i.~e. in a regime where no bound
state exists in the two-body problem ($U/J>-7.9$). Pair formation in the
many-body regime is expected to start at a value of $U/J\approx-2$
\cite{Santos94}.

For strong attractive interactions, i.e. $U/J<-7.9$, the number of doubly
occupied sites saturates. This is in accordance with the fact that the pairs
are well localized on single lattice sites and can be regarded as hardcore
bosons. An increase in molecular fraction due to an attraction-induced
shrinking of the cloud, which would result in a higher average density, is
not substantiated by the following measurements: When tuning the interaction
strengths we could not detect a change in the size of the trapped atom cloud
with our measurement accuracy of 10\%. Furthermore, the same increase in
molecular fraction is found if the attractive interaction is turned on
within only one tunneling time. This demonstrates that we observe local
pairing rather than a redistribution of the trapped atoms on a larger scale.

In conclusion, we have found that the transport of an attractively
interacting Fermi gas in a 3D optical lattice is strongly influenced by the
formation of local pairs. In the future, studying the oscillation frequency
below the superfluid transition temperature could serve to characterize the
BCS-BEC crossover \cite{Wouters2004}. Extending our studies to the repulsive
Fermi-Hubbard model may provide a tool to identify quantum phases such as
the fermionic Mott insulator \cite{Liu2005}.

We would like to thank E.~Altman, G.~Orso, L.~Pollet, M.~Sigrist and F.~Werner
for insightful discussions, and OLAQUI and SNF for funding. Y.~T.
acknowledges support from JSPS and the Huber-Kudlich-Stiftung. This
research was supported in part by the National Science Foundation
under Grant No.~PHY05-51164.

\end{document}